\documentclass[preprint]{revtex4}  
\usepackage{amsmath}
\usepackage{amssymb}
\usepackage{graphicx}
\usepackage[dvips]{epsfig}
\usepackage[ansinew]{inputenc}
\usepackage[T1]{fontenc}
\usepackage{lmodern}
\usepackage{color}

\def\gu{\;{\lower0.3ex\hbox{$\buildrel > \over {\scriptstyle \sim}$}}\;}
\def\lu{\;{\lower0.3ex\hbox{$\buildrel < \over {\scriptstyle \sim}$}}\;}

\begin{document}

\title{Electromagnetic Effects in Capacitively Coupled Plasmas}
\author{D. Eremin}
\affiliation{Institute for Theoretical Electrical Engineering, Ruhr University Bochum, Center for Plasma Science and Technology, D-44780 Bochum, Germany}
\date{\today}
\begin{abstract}
Following requirements of the plasma processing industry for increasing throughput, 
capacitively coupled plasma reactors with large area electrodes driving by very high
frequency sources have been proposed. However, such reactors with plasmas inside
support modes which can negatively influence the uniformity in the ion fluxes or
the average energy of the ions impinging on the substrates, which is an
essential requirement of the industry. It is shown when the popular electrostatic
approximation used for description of the fields in capacitively coupled plasmas (CCP)
breaks down and when these modes must be treated electromagnetically.
Influence of the modes on the essential parameters of the CCP discharges is
discussed. A few techniques for avoiding excitation of the modes
leading to the undesired nonuniformities are mentioned. Results of several
experiments studying such plasma discharges are briefly reviewed.


\end{abstract}

\maketitle

\section{Introduction}

All techincal plasmas, i.e. low-temperature partially ionized plasmas used in the industry for technological purposes such as Plasma Enhanced
Chemical Vapor Deposition (PECVD), etching, etc., are intrinsically bound plasmas generated in corresponding plasma reactors. Unlike
in magnetic fusion plasma reactors, where plasma is too hot to be allowed a contact with the reactor walls without significant consequences and
thus where strong magnetic fields  are used to confine such plasmas, in technical plasmas a direct contact with the reactor electrodes or walls is very common.
In the latter case the electrons, having much greater mobilities compared to that of ions, tend to leave to the metal walls forming a positively
charged sheath due to the ions. This process continues until the potential corresponding to the space charge in the sheaths, which accelerates
the ions and deccelerates the electrons, eventually balances the electron and ion fluxes. Furthermore, the direct contact of plasma with the substrate to be 
processed is often essential, e.g.,  in etching, where ions emanating from the bulk plasma gain energy and directed momentum in passing the sheath region and hit 
the substrate. These ion directed momentum and energy are vital for creating large-aspect structures on the substrate. The ion flux to 
substrate plays also a significant role since it determines rate of the surface processes which occur under the ion bombardment. For a weakly collisional
sheath, the ion flux is proportional to the plasma density at the plasma bulk-sheath interface, which is proportional to the power deposition rate for
high neutral gas pressure (in case of the low pressure energy transport is non-local and this dependence is more complex).
Indeed, to sustain technical plasmas, one needs to couple energy to it, which usually occurs through electromagnetic fields of different frequencies. For the
densities and the field magntidues common to the technical plasmas, energy is coupled predominantly to the electrons. Thiis is due to the much smaller electron mass
in comparison to that of the background neutrals, which results in extremely slow energy exchange between the electrons and the neutral gas, so that 
electron temperatures reach values more than hundred times higher that ion temperatures in the stationary discharges. In contrast, ions and neutrals have comparable masses
and any ion kinetic energy surplus is very efficiently transferred to the background neutral gas, so that the ions end up having practically room temperature. 
The hot electrons ionize the neutrals either through direct impact ionization or through indirect ionization channels, i.e. excitation of metastable atoms or molecules
followed by Penning ionization. Therefore, plasma density is proportional to the reaction rates of such reactions, the reaction rates depend on the electron temperature, and the
electron temperature is affected by the power absorbed by the electrons. 

As the Lorentz force caused by the magnetic field does not perform any work on the charged particles, only electric field can heat plasma. In order to deposit
electric energy into plasma, one first has to excite a certain pattern of electromagnetic fields called mode, which is governed simultaneously by the
Maxwells equations, plasma characteristics and the boundary conditions, including the excitation source. The electromagnetic mode can either exhibit fields with large intensity 
in the plasma bulk or fields more concentrated at the periphery of the discharge, such as sheath-bulk plasma interface. Hence, the former are called
body (or global) modes and the latter surface waves, respectively. 


Once excited, such a mode will dissipate its electromagnetic energy
to the plasma particles through various processes. Common to the technical plasmas are the collisional mode damping due to electron-neutral elastic collisions
and collisionless Landau damping. The collisional damping trasforms the energy of the ordered particle motion under influence of the electromagnetic
forces caused by the mode fields into chaotic thermal motion through charged-neutral particle collisions in case of
low charged particle densities or through Coulomb collisions otherwise. Landau damping
occurs through resonant wave-particle interaction whereby the electromagnetic wave affects mainly the particles trapped in the potential well
of the mode by accelerating the particles with velocities smaller and deccelerating the particles with velocities bigger than the mode phase velocity. 
If the particle velocity distribution function features negative slope at the mode phase velocity, the mode accelerates more particles than it deccelerates
and the effective energy transfer occurs from the mode to the particles. 

For the industrial plasma processing, it is often important to ensure uniformity of the ion fluxes and ion energy distribution throughout the part of the discharge which is relevant to 
the processing. Following the argumentation above, the uniformity of the ion fluxes depends on plasma density,  which is influenced by the absorbed power profile 
(which in turn depends on the distribution of electromagnetic fields in a mode and a specific dissipation process), and energy transport (which may be local in 
case of high collisionality and non-local if the effective mean free path for an energy changing collision is larger than characteristic reactor dimensions).
The uniformity of the characteristic energy of ions impinging on the substrate is directly affected by uniformity of the DC voltage across the sheath (or uniformity
of the DC plasma potential and the DC voltage on the substrate). Therefore, it is important for description of technical plasmas to understand structure of the main 
electromagnetic modes existing in the corresponding reactors, their dissipation, and distribution of the voltage in the plasma sheath.

The simplest technical plasmas are unmagnetized. In this case, the cold collisionless plasma dispersion relation for an unbounded plasma allows propagation of a longitudinal
mode (so that electric field polarization is aligned with the direction of wave propagation) with dispersion relation $\omega = \omega_{pe}$ and a transversal mode
(with the electric field polarization perpendicular to the direction of wave propagation) with dispersion relation $\omega = (\omega_{pe}^2 + k^2c^2)^{1/2}$. 
Hence, in this approximation no body modes are allowed to propagate with frequencies below the electron plasma frequency
(therefore, it is called cutoff frequency). Any wave driven at frequency below the cutoff frequency will be evanescent (heavily damped).
Two different approaches to excitation of the electromagnetic modes (or coupling of the electromagnetic fields
from external sources to plasma) by a time-dependent source can be considered for the technical unmagnetized plasmas, 
which differ by the orientation of the dominant electric field with respect to the plasma surface. For the capacitive (inductive) 
coupling electric field is perpendicular (parallel) to the plasma surface, respectively. The perpendicular electric field in a capacitively coupled plasma discharge
decays into the plasma on a distance of the order of the Debye length to a small value needed to sustain the Ohmic current ensuring the current continuity. 
However, the Maxwell's equations require that the parallel magnetic field generated by the time-changing perpendicular electric field must decay into the plasma
on a distance of order of the inertial skin depth $\delta_i \equiv c/\omega_{pe}$, and the decaying magnetic field generates a parallell electric field. Thus,
there is no solution with a purely perpendicular electric field, there is always a radial component of the electric field which is small under normal
conditions in a capacitively coupled discharge. As will be demonstrated below, in the cold plasma approximation a capacitively coupled  
discharge support two different dominant modes when an analytical treatment takes into account the bounded nature of such a discharge, 
calculating the fields in each of the separate discharge regions (plasma sheaths and bulk) respecting the boundary conditions and matching
the solutions on the interfaces between the different regions. These are normal modes of plasma filled cavity representing a capacitively
coupled reactor.

These modes differ by their symmetry of the perpendicular electric field with respect to the discharge midplane (and hence are called "even" and "odd" modes)
and lead to different physical consequences. If $\delta_i$ or its collisional analog (in case of the collision dominated skin effect) $\delta_c = \delta_i(2\nu / \omega)^{1/2}$
are smaller than the bulk plasma width $d$, the fields tend to be concentrated around the sheath-bulk plasma interface. Therefore, these
modes in a capacitively coupled discharge consist of surface waves. If the mode wavelengths ($\lambda_e \approx \lambda_0 (s/d)^{1/2}$ and
$\lambda_o \approx \lambda_0 (sd)^{1/2}/\delta$ for the even and odd modes, respectively. Here  $\lambda_0=2\pi c/\omega$ is the vacuum wavelength for the frequency $\omega$ and
$\delta=max(\delta_i,\delta_c)$) become comparable to the electrode radius $R$, the wave nature of the modes start to manifest itself
through the nonuniform distribution of the discharge voltage ("standing wave effect") or plasma potential ("telegraph effect") in the radial direction. 
It will be shown below that a strong skin effect can lead to a redistribution of the fields inside a mode, making the fields stronger towards the
discharge edge. The action of the skin effect is akin to that observed in metals conducting a radio-frequency (RF) current, where the skin effect leads to
generation of eddy currents reducing the original current in the bulk of the conductor and increasing it at the periphery of the conductor, so
that effectively the RF current tends to be concentrated close to the periphery of the discharge. Likewise, plasma in a cylindrical capacitively coupled
reactor the even mode shows increase of the dominant axial current towards the edge and the odd mode shows increase of the dominant radial
current towards the plasma sheaths. Nevertheless, the field symmetry remains unchanged even when strong skin effect is present, 
so that classification of the modes remain the same.
It is important to note that these modes are so-called slow modes, i.e. their phase velocity is smaller than the speed of light and thus they can be dissipated not
only through collisions but also through the collisionless Landau damping. 

Usually,  in the capacitively coupled discharges $\delta_i$ is large compared to the plasma bulk size, thus the magnetic field hardly changes in the plasma
and the parallel electric field is negligible, and $min(\lambda_e,\lambda_o)\gg R$ so that the wave nature of the modes is hardly noticeable. 
In this case one can adequately describe the modes using an electrostatic approximation, which uses solely Poisson equation to solve for the electric field.  
The discharge voltage at the driven electrode and the plasma potential are both uniform in the radial direction. However, there is a growing trend to increase the 
substrate size and driving frequency of such discharges in order to increase the processing throughput. In the light of the above arguments, under
these circumstances the field pattern might change considerably and lead to large nonuniformities in the power depositions and sheath voltage drops.
Furthermore, a strong skin effect (when $\delta$ is small because of the higher plasma densities) can lead to a change in the type of the power deposition, changing it from capacitive
type when the perpendicular electric field dominates to the inductive type, when the radial electric field dominates.  
This demands an electromagnetic description despite the vacuum wavelength corresponding to the driving frequency is commonly greater than the reactor 
dimensions.

The inductively coupled discharges (ICPs) use predominantly parallel electric field for power deposition into the plasma. As the radial electric field
for the overdense plasmas, which is normally the case, decays into the plasma with the skin-effect,
the power absorption can be more or less uniform (when the skin depth is large compared to the interelectrode gap) or strongly localized otherwise.
In principle, as the plasma is usually overdense, such system supports similar modes to the capacitively coupled discharges. As the large
parallel electric in the inductively coupled discharges field penetrates plasma much deeper than the perpendicular electric field in the capacitively
coupled discharge in their normal regimes of operation, energy coupling is more efficient in case of the inductively coupled discharges, which
leads to higher plasma densities attained in such discharges. As in capacitively coupled discharges which exhibit transit to the inductively 
coupled regime under extreme conditions, the inductively coupled discharges can be also operated in the capacitively coupled regime.

If the magnetic field is used in the discharge, the magnetized plasma dispersion relation allows propagation of the body helicon modes with 
frequencies well below electron plasma frequency.
Since these modes are global, they allow more effective and uniform energy coupling between antenna and plasma.  

Whereas the electromagnetic effects are well described for the plasma sources whose operation depends on those very effects (e.g., see
textbooks \cite{lieberman_2005} and \cite{chabert_2011}), it
has been only relatively recently realized that the electromagnetic effects become significant for the modern capacitively coupled discharges
with large area electrodes driven by high frequency sources. Here, the attention will be paid mostly to description of the electromagnetic effects 
for the modes in a capacively coupled reactor.

\section{Surface wave modes supported by a cylindrical capacitively coupled plasma reactor}\label{s1}

Many physical insights can be obtained by considering a simple cylindrical cavity
representing capactiviely coupled reactors filled with plasma modelled as
a dielectric using the cold homogeneous unmagnetized plasma model, the plasma being
separated from the electrodes by sheaths.
In description of such a model we will mostly follow \cite{sansonnens_2006}. 

\begin{figure}[ht]
\centering
\epsfig{file=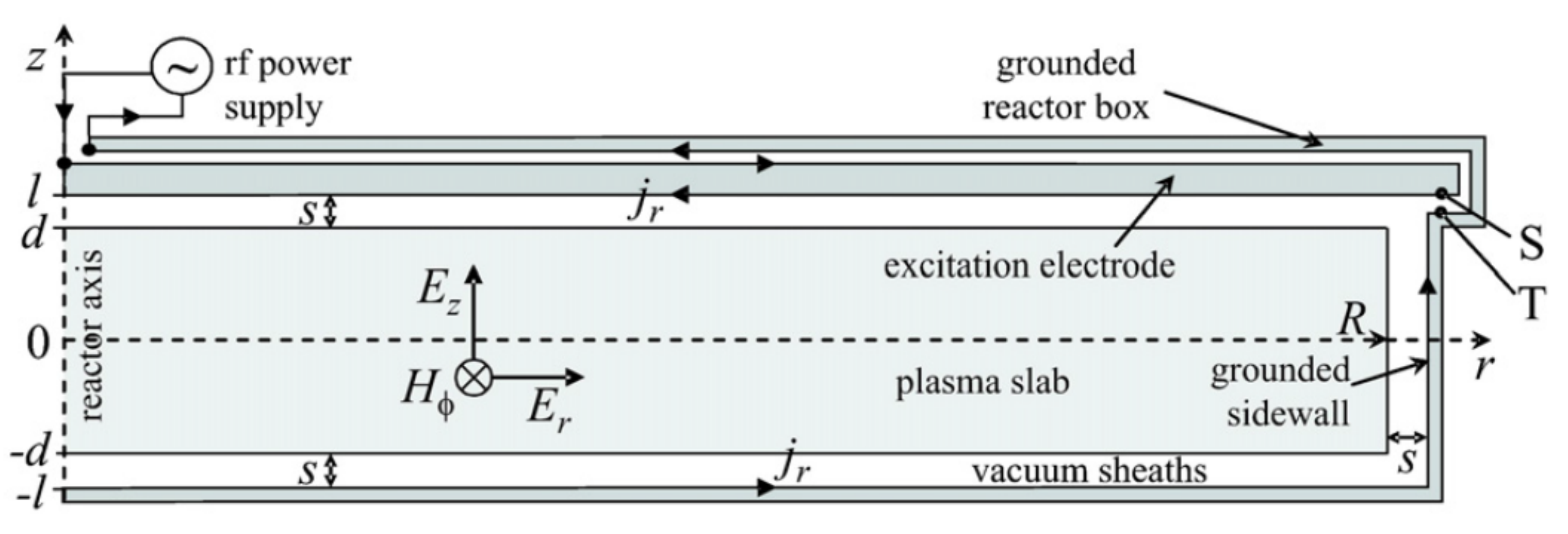,width=12cm}
\vspace{0.3cm}
\caption{Geometry of a typical asymmetric capacitively coupled reactor, taken from \cite{howling_2007}.}
\label{fig1}
\end{figure}

A capacitively coupled reactor usually consists of two planar electrodes, of which at least
one is driven by an external AC source, the driving frequency being in the range of
RF frequencies. Most of the voltage drop takes place in the positively charged sheaths, which
separate bulk quasineutral plasma from the electrodes. For the lateral
confinement a sidewall made of metal or dielectric material is usually used. A typical 
CCP reactor with cylindrical geometry is depicted in Fig.~\ref{fig1}. Since the skin depth in metal for RF frequencies is much smaller than the electrode
thickness, the supplied voltage propagates first in the form of a TEM mode in the waveguide formed by the
back side of the driven electrode and the grounded encasing from the point of application of the external
source to the slit at the radial periphery of the reactor (denoted 'ST' in the figure), at which point electromagnetic waves supplying power
for sustaining the plasma enter the reactor and carry the energy further in the discharge in the form
of the modes supported by such configuration from the radial periphery to the radial center
of the discharge. Note that the CCP reactor shown in Fig.~\ref{fig1} has asymmetric configuration, i.e.
the total area of the grounded surfaces is larger than the driven electrode area. Such
a configuration is more general than a symmetric one as it allows excitation of the
both main modes supported by the CCP reactors with plasma, as will be shown shortly.

Without plasma, such reactor is a simple capacitor, which for the usual frequencies
and sizes support only TEM modes propagating with the speed of light, as these modes have no cutoff frequency from
below. The first available TM mode in such a waveguide cannot propagate for $l<\lambda_0/4$
with $l$ the interelectrode distance and $\lambda_0 = c/\omega$ the vacuum wavelength \cite{ramo}. 

A presence of plasma dramatically changes the modes supported by a capacitively coupled
reactor. The electrode inductance for the radial currents flowing through the electrodes
is not altered by the plasma, but the capacitance increases by a factor of $l/s$
compared to the vacuum case because plasma virtually screens the fields from it, hence
the capacitance is determined by the sheaths only. The mode becomes a slow wave,
its phase velocity $v_{ph} = \omega/k = c (s/l)^{1/2}$ (with $k$ the radial wave vector) being significantly smaller than the speed of
light owing to $s\ll l$ \cite{howling_2007}. Therefore, the effective radial wavelength of such a mode $\lambda=2\pi/k$ is 
much smaller than wavelength of the vacuum TEM modes $\lambda_0$ and can become comparable to the reactor radius (so
that the mode's wave nature leads to signficant nonuniformities) at  much smaller frequency than expected from an estimate based on the vacuum TEM modes. 

To analyze such model a convenient choice (e.g., \cite{bowers_2001}) is to represent the plasma by a uniform ion
background and electrons as a central lossy dielectric slab with permittivity equal to that of a cold 
collisional plasma, $\epsilon_p(\omega) = 1 - \frac{\omega_{pe}^2}{\omega(\omega-i\nu)}$ with
$\nu$ the frequency of electron-neutral elastic collisions and $\omega_{pe}^2 = e^2n_{e}/(\epsilon_0 m_e)$
the plasma density where electron density $n_e$ is assumed homogeneous. The central electron slab has the same
density as that of the ion background, so that the bulk plasma slab is quasineutral and the matrix sheaths
are positively charged. As has been argued in the Appendix~A of \cite{bowers_2001}, linear small
amplitude oscillations of the electron negatively charged slab around its stationary position
can be adequately described by a stationary dielectric slab because the charge perturbation 
resulting from the slab motion is mimicked by the harmonic charges occuring at the sheath-dielectric
interface in order to satisfy boundary condition for the perpendicular electric flux density.
All sheaths have uniform thickness $s$.
Although such a model is clearly not self-consistent and does not account for many potentially significant effects such as, among others,
non-uniform plasma profile, finite temperature effects, nonlinear sheath oscillations, and nonuniform
sheath thickness, it provides basic
ideas about the modes which are supported by a CCP reactor. 

Futher, we will assume the field pattern of a TM mode (e.g., \cite{bowers_2001}, \cite{lieberman_2002}), i.e. the electric field with the radial and axial
components and the magnetic field with the azimuthal component. Assuming the azimuthal symmetry,
all the fields depend only on the radial and the axial coordinates. As discussed above, there is always
a radial component of the electric field because the Maxwell's equations predict decay of the magnetic
field into the plasma and a spatially changing magnetic field is accompanied by a radial electric field component. 
However, when the driving frequency and electrode radius of a CCP reactor are sufficiently small so that the
nonuniformity of the fields is insignificant (which had been usually the case up to relatively recently)
, the radial electric field is negligibly small compared to the axial field, some authors prefer to call 
such modes quasi-TEM (e.g., \cite{sansonnens_2006}). In \cite{bowers_2001} it was shown that
the modes with the TE-like field pattern have phase velocities greater than the speed of light so
that their wavelengths are much bigger than those of the TM-like modes, thus the TE-like modes are not
considered here. 

Following the ansatz, the Ampere's and Faraday's laws can be reduced to a single equation for the azimuthal component
of the magnetic field. When written in the sheaths and bulk plasma separately, it reads 
\begin{equation}
\frac{1}{\epsilon}\left[ 
\frac{\partial^2 H_{\phi}}{\partial r^2} + \frac{1}{r}\frac{\partial H_{\phi}}{\partial r} - \frac{1}{r^2} H_{\phi} 
\right] 
+ \frac{\partial}{\partial z}\frac{1}{\epsilon}\frac{\partial H_{\phi}}{\partial z} + k_0^2 H_\phi = 0,
\label{eq1}
\end{equation} 
where $\epsilon=1$ in the sheaths and $\epsilon=\epsilon_p$ in the bulk plasma and $k_0=\omega/c$. 
This equation has to be solved under boundary condition of vanishing tangential electric field at the
metal electrodes, which translates to $\partial H_\phi / \partial z = 0$ according to the Ampere's law. 

Then, assuming that all fields change in time harmonically with frequency $\omega$
and solving this equation for $H_\phi$ using the method of separation of variables
along with calculating the axial and the radial electric field components from the
Ampere's law yields (see Ref. \cite{sansonnens_2006}) for the bulk plasma ($-d \le z \le d$)
\begin{equation}
\begin{array}{l}
H_\phi = \sum\limits^\infty_{n=0} \frac{C_n^e}{K_{zp,n}^e}\cos(K^e_{zp,n} z)J_1(K^e_{r,n} r) +  \sum\limits^\infty_{n=0} \frac{C_n^o}{K_{zp,n}^o}\sin(K^o_{zp,n} z)J_1(K^o_{r,n} r) \\
E_z = \frac{1}{j\omega\epsilon_0\epsilon_p}\left(\sum\limits^\infty_{n=0} \frac{C_n^eK^e_{r,n}}{K_{zp,n}^e}\cos(K^e_{zp,n} z)J_0(K^e_{r,n} r) +  \sum\limits^\infty_{n=0} \frac{C_n^oK^o_{r,n}}{K_{zp,n}^o}\sin(K^o_{zp,n} z)J_0(K^o_{r,n} r) \right) \\ 
E_r = \frac{1}{j\omega\epsilon_0\epsilon_p}\left( \sum\limits^\infty_{n=0} C_n^e\sin(K^e_{zp,n} z)J_1(K^e_{r,n} r) -  \sum\limits^\infty_{n=0}  C_n^o\cos(K^o_{zp,n} z)J_1(K^o_{r,n} r) \right)
\end{array} \label{eq2}
\end{equation} 
and for the sheath plasma $d \le \pm z \le l$
\begin{equation}
\begin{array}{l}
H_\phi = \sum\limits^\infty_{n=0} \frac{D_n^e}{K_{zs,n}^e}\cos(K^e_{zs,n} (l\mp z))J_1(K^e_{r,n} r) \pm  \sum\limits^\infty_{n=0} \frac{D_n^o}{K_{zs,n}^o}\cos(K^o_{zs,n} (l\mp z))J_1(K^o_{r,n} r) \\
E_z = \frac{1}{j\omega\epsilon_0}\left(\sum\limits^\infty_{n=0} \frac{D_n^eK^e_{r,n}}{K_{zs,n}^e}\cos(K^e_{zs,n} (l\mp z))J_0(K^e_{r,n} r) 
\pm  \sum\limits^\infty_{n=0} \frac{D_n^oK^o_{r,n}}{K_{zs,n}^o}\cos(K^o_{zs,n} (l \pm z))J_0(K^o_{r,n} r) \right) \\ 
E_r = \frac{1}{j\omega\epsilon_0}\left( \mp \sum\limits^\infty_{n=0} D_n^e\sin(K^e_{zs,n} (l\mp z))J_1(K^e_{r,n} r) -  \sum\limits^\infty_{n=0}  D_n^o\sin(K^o_{zs,n} (l\mp z))J_1(K^o_{r,n} r) \right) ,
\end{array}  \label{eq3}
\end{equation}
where $J_0$ and $J_1$ are zeroth and first order Bessel functions of the first kind representing the standing waves formed by the counterpropagating
surface wave modes, $C$ and $D$ are mode amplitudes, $K_r$ and $K_z$ are complex wavevectors coupled by
\begin{equation}
K_{zp,n}^2 = k_0^2\epsilon_p - K_{r,n}^2  \label{eq4}
\end{equation}
and
\begin{equation}
K_{zs,n}^2 = k_0^2 - K_{r,n}^2  \label{eq5}
\end{equation}
resulting from substitution of Eqs.~(\ref{eq4}) and (\ref{eq5}) into Eq.~(\ref{eq1}).
All the modes are classified by their symmetry of $H_\phi$ (or of $E_z$, which has the same symmetry) with respect to $z$ (the even and odd modes being denoted with the superscript 'e' and 'o',
respectively) and by their axial mode number $n$. It will be demonstrated shortly that the $n=0$ modes are dominant and the Tonks-Dattner-like surface modes with $n\ge 1$
are damped heavily in the radial direction and are non-negligible only close to the lateral sidewalls, where they are needed to satisfy boundary conditions 
and cause strong local field nonuniformities \cite{lieberman_2002}. Fig.~2 shows characteristic field patterns for the first even and odd modes.

\begin{figure}[ht]
\centering
\epsfig{file=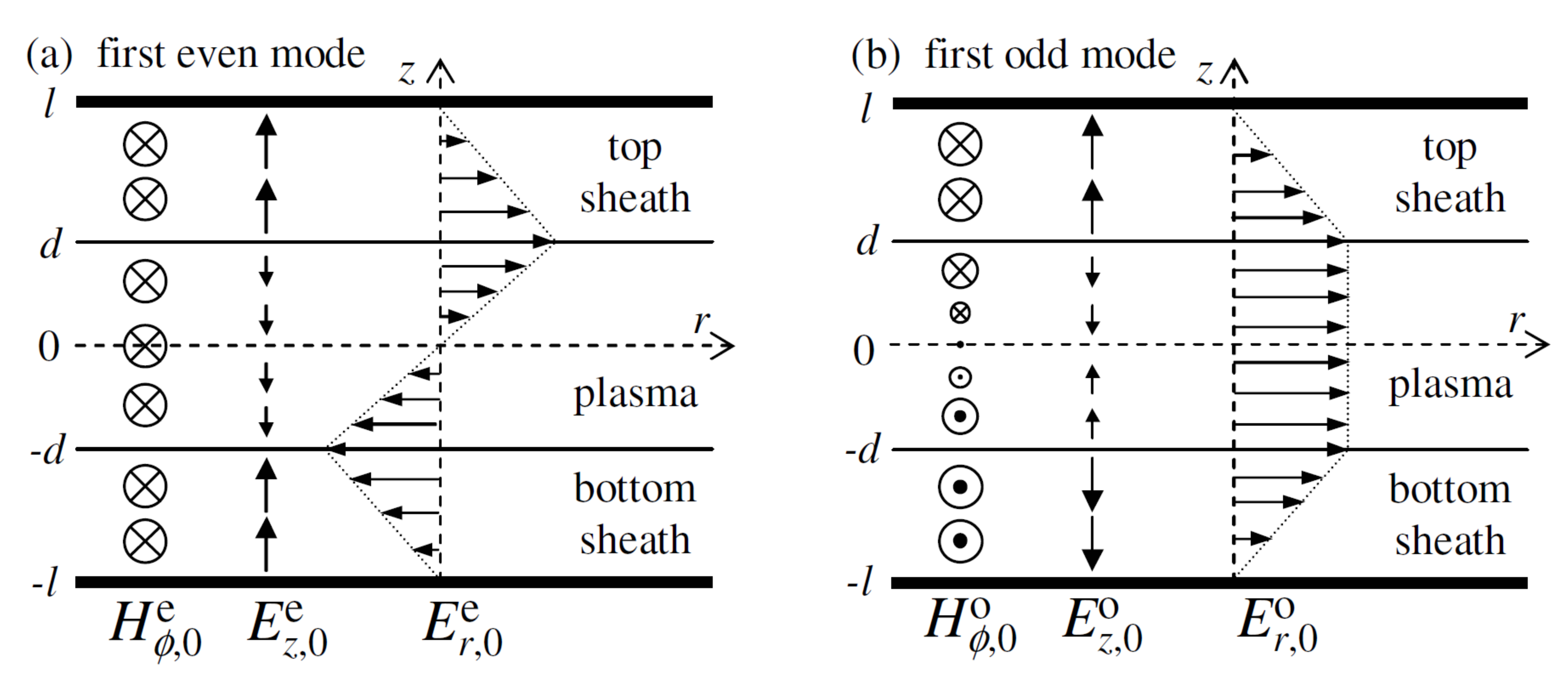,width=12cm}
\vspace{0.3cm}
\caption{Characteristic field patterns of the first even and odd modes in a CCP reactor, taken from \cite{sansonnens_2006}.}
\label{fig2}
\end{figure}

To find the dispersion relations of the even and the odd modes, the matching conditions at the interface between the sheaths and the bulk plasma are used. The
radial electric field component $E_r$  and the axial electric flux density $\epsilon E_z$  must be continuous there. It is worth noting that the 
continuity of the axial electric flux density at the interface between the sheaths and the bulk plasma also means discontinuity of the
axial electric field there, which points out at the accumulation of surface charge. As has been mentioned above (see also \cite{bowers_2001}), this surface 
charge mimicks charge perturbation due to the electron plasma slab motion in a more realistic treatment. The first matching
condition at $z=d$ for Eqs.~(\ref{eq2}) and (\ref{eq3}) provides the relation between the mode amplitudes in the sheath and the bulk plasma, and
the second matching condition yields the dispersion relation for the even modes,
\begin{equation}
\epsilon_p K_{zs,n}^e \sin (K_{zs,n}^e s) \cos(K_{zs,p}^e d) + K_{zp,n}^e \cos (K_{zs,n}^e s) \sin(K_{zs,p}^e d) = 0 \label{eq6}
\end{equation}
and for the odd modes,
\begin{equation}
\epsilon_p K_{zs,n}^o \sin (K_{zs,n}^o s) \sin(K_{zs,p}^o d) - K_{zp,n}^o \cos (K_{zs,n}^o s) \cos(K_{zs,p}^o d) = 0. \label{eq7}
\end{equation}

\subsection*{Long radial wavelength asymptote} \label{subsection_long_wavelength}

It is useful to examine the asymptotes with respect to real $K_r$  of the dispersion relations (\ref{eq6}) and (\ref{eq7}) for the $n=0$ modes . 
For the long wavelengths $K_{r,0} \to 0$, assuming also that $\omega \to 0$ (this assumption will be verified afterward), 
$|K_{zs,0}|s \ll 1$, and $K_{zp,0} = j/\delta_s$
(where $\delta_s = c/(\omega\sqrt{1-\epsilon_p})$ defines generalized complex skin depth, which
in the collisionless limit $\nu\ll \omega$ can be approximated by the inertial skin depth and
can be approximated as $\delta_c/(1-j)$ with $\delta_c = \delta_i(2\nu/\omega)^{1/2}$ when
the collisions are strong, $\nu\gg \omega$; anomalous skin effect is not considered here),
using expansion of the trigonometric functions for a small argument, utilizing Eqs.~(\ref{eq4})
and (\ref{eq5}), one obtains
\begin{equation}
\frac{\omega}{K_{r,0}^e} = \frac{c}{\left(1+\frac{\delta_s}{s}\tanh\frac{d}{\delta_s}\right)^{1/2}} \label{eq8}
\end{equation}
for the even mode and 
\begin{equation}
\frac{\omega}{K_{r,0}^o} = \frac{c}{\left(1+\frac{\delta_s}{s}\coth\frac{d}{\delta_s}\right)^{1/2}} \label{eq9}
\end{equation}
for the odd mode (see also \cite{bowers_2001}). Note that the phase velocities of both modes are smaller than the speed
of light and that the odd mode is slower than the even mode. In both cases $\omega \to 0$ as $K_{r,0}\to 0$ 
in accord with the assumption made in the beginning of the derivation. 

If $\delta_s\to 0$ than the bulk plasma is similar to a metal conductor
and effectively screens the
fields entering it and the coupling between the surface waves propagating in the top and the bottom sheaths is
strongly suppressed. In this case Eqs.~(\ref{eq8}) and (\ref{eq9}) transform into dispersion relation of TEM modes $\omega = K_{r,0} c$
propagating in the waveguides formed in the sheaths bounded by the electrode on one side and the bulk plasma on the other
(see also \cite{chabert_2011}). 

\subsection*{Short radial wavelength asymptote} \label{subsection_short_wavelength}

Assuming $K_{r,0} \to \infty$, from Eqs.~(\ref{eq4}) and (\ref{eq5}) it follows that $K_{zs,0} = i K_{r,0}$ and $K_{zp,0} = i K_{r,0}$.
From Eqs.~(\ref{eq6}) and (\ref{eq7}) it follows that in this limit the dispersion for the first even and odd modes becomes
\begin{equation}
\omega = \frac{\omega_{pe}}{\sqrt{2}}, \label{eq10}
\end{equation}
the factor of $\sqrt{2}$ being due to the bounded geometry of the discharge. The same short wave asymptotic behavior is demonstrated,
for example,
for the Travelpiece-Gould surface mode propagating in a cylindircal plasma column (e.g., \cite{swanson_2003}).

\subsection*{Higher oder modes}

The dispersion relations for the higher oder modes $n\ge 1$ can obtained by letting the first term in Eq.~(\ref{eq6}) vanish 
(it is assumed that for the higher order modes the first term  in Eq.~(\ref{eq6}) is much greater than the second one due to $|\epsilon_p|\gg 1$ 
for the typical parameters of CCP discharges, an assumption which can be verified \emph{a posteriori}; note that for the first mode the two terms are
assumed to be comparable). This provides two types of higher oder modes for the even and odd symmetry. The first type of the
higher order even modes has 
\begin{equation}
K^e_{zs,n} = \frac{n\pi}{s}, \label{eq9-1}
\end{equation}
and the second type has
\begin{equation}
K^e_{zp,n} = \frac{(2n-1)\pi}{2d}. \label{eq9-2}
\end{equation}
One can see that the first type of modes with the sheath axial wavevector given in Eq.~(\ref{eq9-1}) arise
due to formation of the standing waves in the sheath region, similar to the Tonks-Dattner resonance
modes. For the latter modes however, different resonances occur due to the dispersion of 
the Langmuir plasma waves caused by thermal effects (e.g., \cite{crawford_1963}), in the case investigated here
the dispersion occurs due to the bounded nature of the CCP reactor. The second type of modes with 
the bulk plasma axial wavevector provided by Eq.~(\ref{eq9-2}) is generated by the standing waves formed in the bulk plasma.
For the first mode type Eq.~(\ref{eq5}) gives the dispersion relation 
\begin{equation} 
\omega = c\left(\frac{n^2\pi^2}{s^2}+(K_{r,n}^e)^2\right)^{1/2} \label{eq9-3}
\end{equation}
Similarly, for the second mode type Eq.~(\ref{eq4}) yields
\begin{equation} 
\omega = c\left(\frac{1}{\delta_s^2}+\frac{(2n-1)^2\pi^2}{4d^2}+(K_{r,n}^e)^2\right)^{1/2} \label{eq9-4}
\end{equation}
Analogously, there are two similar mode types for the higher order odd modes. Applying the same
procedure to Eq.~(\ref{eq6}), one obtains for the first type of 
the higher order odd modes 
\begin{equation}
K^o_{zs,n} = \frac{n\pi}{s}, \label{eq9-5}
\end{equation}
and for the second type
\begin{equation}
K^o_{zp,n} = \frac{n \pi}{d}. \label{eq9-6}
\end{equation}
This yields dispersion relation for the first type of the higher order odd modes as
\begin{equation} 
\omega = c\left(\frac{n^2\pi^2}{s^2}+(K_{r,n}^e)^2\right)^{1/2}, \label{eq9-7}
\end{equation}
and for the second type
\begin{equation} 
\omega = c\left(\frac{1}{\delta_s^2}+\frac{n^2\pi^2}{d^2}+(K_{r,n}^e)^2\right)^{1/2}. \label{eq9-8}
\end{equation}

\subsection*{Electrostatic limit}

It is also interesting to consider the electrostatic limit for the dispersion relations of the first even and the odd modes to see
when the electrostatic approximation provides a reasonable approximation to the EM solution, and when it breaks down.
For simplicity only collisionless case is considered. The electrostatic approximation assumes instantaneous change of the electrostatic field
after a change in the source charges, in the entire space. By
letting the speed of light $c$ in Eqs.~(\ref{eq6}) and (\ref{eq7}) to go to infinity, in this case one can obtain exact expressions 
for the dispersion relations of the first even and odd modes
\begin{equation}
\omega = \omega_{pe} \left(\frac{\coth (K_{r,0}^e d)}{\coth(K_{r,0}^e s) + \coth(K_{r,0}^e d)}\right)^{1/2} \label{eq11}
\end{equation}
and
\begin{equation}
\omega = \omega_{pe} \left(\frac{\tanh (K_{r,0}^o d)}{\coth(K_{r,0}^o s) + \tanh(K_{r,0}^o d)}\right)^{1/2}  \label{eq12}
\end{equation}
(see also \cite{bowers_2001} and \cite{cooperberg_1998}). Additionally, there are also trivial solutions 
allowing arbitrary $\omega$ for $K_{r,0}^e=0$ or $K_{r,0}^o=0$.

The long radial wavelength limit of Eqs.~(\ref{eq11}) and (\ref{eq12}) coincide with the electromagnetic case. 
The short radial wavelength limit yields
\begin{equation}
\omega = \omega_{pe}\sqrt{\frac{s}{l}} \label{eq13}
\end{equation}
for the first even mode and
\begin{equation}
\frac{\omega}{k} = \omega_{pe}\sqrt{\frac{ds}{2}} \label{eq14}
\end{equation}
for the first odd mode. Whereas Eq.~(\ref{eq14}) is close to the EM analog (the small difference is
due to the field nonuniformities caused by the skin effect), Eq.~(\ref{eq9}) in the collisionless
case with a large skin depth compared to the bulk plasma width, $d/\delta_i \ll 1$, Eq.~(\ref{eq13}) is drastically different
from the EM case, Eq.~(\ref{eq9}) in that it predicts cutoff of the even mode propagation at the electron series resonance
frequency. This means that in the framework of the electrostatic analysis the electrostatic field with the even symmetry
(not that such modes are the only modes existing in a geometrically symmetric reactor), driven by a
power source at a frequency below the electron series resonance, is distributed not in the form of the electrostatic
analog of the even surface wave mode, but according to the trivial soluition predicting uniform voltage distribution
over the driven electrode (such field pattern is a very popular assumption in modelling the CCP discharges. As one can
see, it has limited validity). Therefore, there is a knee around the elecron series resonance frequency
 in the long wavelength part of both electrostatic and electromagnetic dispersion relations
for the first even surface mode (see Fig.~3). In the electrostatic approximation this knee is sharp, in the electromagnetic
treatment it is much more rounded. One can however see the transition of the smooth dispersion relation in the EM case
to the sharp knee in the electrostatic dispersion relation if one gradually increases the speed of light
(see Fig.~4.4 in \cite{bowers_2001}). 

\begin{figure}[ht]
\centering
\epsfig{file=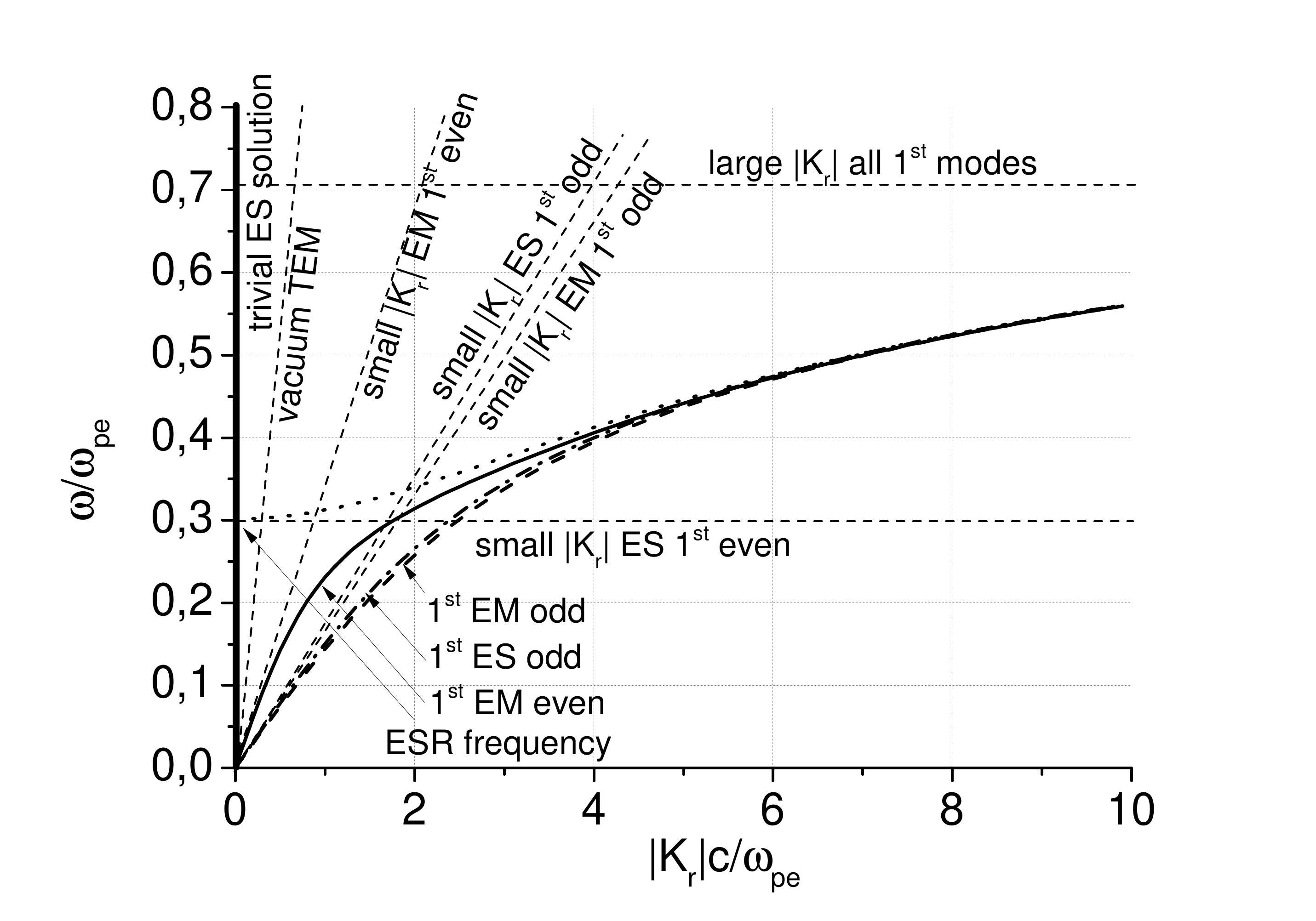,width=12cm}
\vspace{0.3cm}
\caption{Cold plasma dispersion relation. The higher modes have significantly higher frequencies and are not shown}
\label{fig3}
\end{figure}

Therefore, one can conclude that the electromagnetic treatment is important for the description of the first even mode
effects where dispersion relation plays a significant role, for example, when the electrode radius becomes comparable
to the mode wavelength or when the Landau resonant wave-particle interaction at the mode phase velocity is taken into account.
Another situation where the electromagnetic treatment can be important is when plasma density is so large, 
that the skin width becomes comparable to the interelectrode gap. In this case the field nonuniformities
can lead to significant redistribution of the current and power deposition even when the radial size of a 
reactor is small compared to the wavelength of the first even or odd modes \cite{mussenbrock_2008}.
 
The higher order modes ($n \ge 1$) described in the previous sections do not exist in the electrostatic cold plasma approximation
(however, similar modes arise in the electrostatic approximation if one accounts for the thermal effects introducing 
dispersion in the dispersion relation of the electrostatic plasma waves. The latter modes are
called Tonks-Dattner resonance modes). Thus, the EM treatment is essential as long as these modes are concerned.

One can also observe that the electromagnetic and electrostatic dispersion relations for first even and odd modes  
approach each other in the short radial wavelength limit. It can be concluded that in this limit, the corresponding modes
become virtually electrostatic, which is born out by estimation of the electromagnetic and electrostatic
energy content for the corresponding modes \cite{bowers_2001}.

\subsection*{Even modes}
 The first even mode ($n=0$) is usually excited in the long radial wavelength part of the corresponding dispersion relation (see Fig.~\ref{fig3}). 
Following the same assumptions as in the derivation of Eq.~(\ref{eq8}) , one obtains
\begin{equation}
\begin{array}{l}
(K_{r,0}^e)^2 \approx k^2_0 \frac{(1 + d/s)}{(1 + d/(\epsilon_p s))} \to k_0^2(1+ d/s) \\
(K_{zs,0}^e)^2 \approx -k^2_0 \frac{d(1 - 1/\epsilon_p)}{s(1 + d/(\epsilon_p s))} \to  -k_0^2 d/s \\
(K_{zp,0}^e)^2 \approx k^2_0 \frac{d(\epsilon_p - 1)}{s(1 + d/(\epsilon_p s))} \to  k_0^2 (\epsilon_p-1)
\end{array}, \label{eq15}
\end{equation}
where an additional limit of $|\epsilon_p|\gg d/s$ was taken (\cite{lieberman_2002},\cite{sansonnens_2006}). As one can see, $K_{r,0}^e$ is real,
$K_{zs,0}^e$ is imaginary, and $K_{zp,0}^e$ is complex (in the collisionless case it is also purely
imaginary).  This solution describes a standing surface wave mode
formed by the surface waves propagating radially inward from the plasma periphery, where they are excited at
the slit 'ST' (see the Fig.~(\ref{fig1})). The fields decay in the axial dimension, however, this effect is
small if the skin depth is large compared to the interelectrode gap. However, when the skin effect is strong,
the fields tend to concentrate in the vicinity of the interface between the sheath and bulk plasma,
so that the modes become truly surface waves.

It might appear that Eq.~(\ref{eq15}) predicts a resonance at the electron series frequency as the 
denominator $1 + d/(\epsilon_p s)$ vanishes there. However, this leads to $K\to \infty$, which
is inconsistent with the assumption made in the derivation of this equation that $|K_{zs,0}^e|s\ll 1$ and $|K_{zp,0}^e|d \ll 1$. 
A consistent derivation of the long radial wavelength asymptote 
(see Eq.~(\ref{eq8})) does not predict the resonance to occur in the long wavelength limit,
which is also corroborated numerically (see Fig.~3).

\begin{figure}[ht]
\centering
\epsfig{file=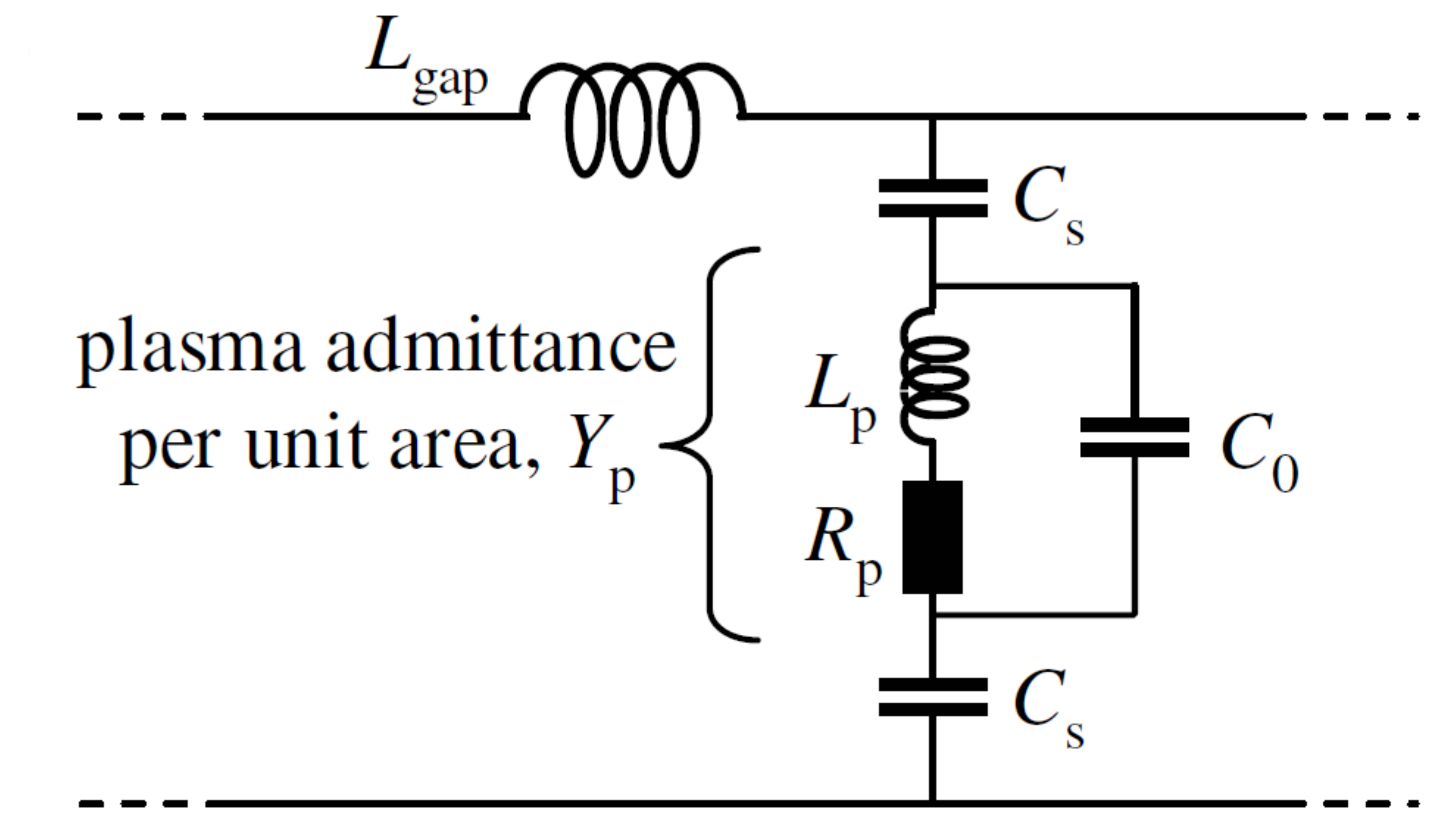,width=7cm}
\vspace{0.3cm}
\caption{Equivalent circuit for description of the first even mode propagation, taken from \cite{sansonnens_2006}.}
\label{fig4}
\end{figure}

It is also helpful for understanding of the process of the mode propagation
to construct an equivalent one-dimensional circuit to describe the propagation of the first even mode based
on the transmission line analogy, see Fig.~\ref{fig4} (see, e.g., \cite{chabert_2004}, \cite{sansonnens_2006}). 
The series impedance per unit area $Z=L_{gap}$ is caused by inductance of the interelectrode gap per unit area
$L_{gap}=2\mu_0 l$ (this value is hardly different from its vacuum analog
provided the skin effect is weak, since the RF current density is still practically uniform in the entire discharge in this case)
, and the parallel admittance per unit area comes from the sheath capacitance
per unit area $C_s = \epsilon_0/s$ and the bulk plasma admittance 
which can be calculated as a capacitance of a capacitor with the interelectrode gap $2d$ filled with a dielectric
having permittivity of $\epsilon_p$, $Y_p= j\omega\epsilon_0\epsilon_p/2d$ \cite{lieberman_1994}.
Substituting the expression for $\epsilon_p$, it can be seen that $Y_p$ can stem from
a bulk vacuum capacitance per unit area $C_0 = \epsilon_0/2d$,  
from bulk plasma (non-magnetic) inductance caused by the electron inertia $L_p = 1/(C_0 \omega_{pe}^2)$, and 
from the bulk plasma resistance caused by the electron-neutral elastic collisions $R_p = \nu L_p$. 

Assuming $I$ to be the current flowing through a unit area and the $V$ the voltage between the electrodes
at a position $x$ ($V(x) \equiv -\int^l_{-l} E_z(z,x) dz$), one can write differential equations for $I$ and $V$
at the transmission line interval located between $x$ and $x+dx$: $\partial V/\partial x = - IZ$ and $\partial I / \partial x = -VY$, respectively,
which reduces to
\begin{equation}
\frac{\partial^2 V}{\partial x^2} = - K^2 V, \label{eq16}
\end{equation}
where $K^2 = -ZY = k_0^2 \frac{(1+d/s)}{(1+d/(\epsilon_p s))}$, which agrees with Eq.~(\ref{eq15}). 

This derivation is based on the assumption of the uniform fields  and also misleadingly predicts a resonance at the
electron series resonance frequency (see the discussion after Eq.~(\ref{eq15})).
This assumption of field uniformity is violated if the skin effect is sufficiently strong.
 In the latter case one can use the long wavelength asympote of the exact field solutions in Eqs.~(\ref{eq2}) and (\ref{eq3})
(similar to the derivation which led to Eq.~(\ref{eq8}))
to deduce more accurate expressions for the lumped series impedance $Z\equiv j\omega L$ and the lumped parallel admittance $Y \equiv j\omega C$ 
per unit length (see \cite{chabert_2007}).  Adopting cylindrical geometry, 
calculating voltage across the electrodes at a radial position $r$
again as $V(r) \equiv -\int^l_{-l} E_z(r,z) dz$ 
and the total current flowing through the electrode area $\pi r^2$ as $I(r) = 2\pi r H_\phi (r,z=l)$,
one can write for the impedance of the line $V/I = \sqrt{L/C}$ . In addition, for the wavevector of the
perturbation propagating along the transmission line one can write $K_{r,0}^e = \omega\sqrt{LC}$, where
$K_{r,0}^e$ is to be calculated from Eq.~(\ref{eq8}). One can then obtain 
\begin{equation}
\begin{array}{l}
L = \mu_0\frac{s}{\pi r}\left(1 + \frac{\delta_s}{s}\tanh\frac{d}{\delta}\right)\left(1 - \frac{\omega^2}{\omega_{pe}^2}\frac{\delta_s}{s}\tanh\frac{d}{\delta_s}\right) \\
C = \epsilon_0 \frac{\pi r}{s} \left(1 - \frac{\omega^2}{\omega_{pe}^2}\frac{\delta_s}{s}\tanh\frac{d}{\delta_s}\right)^{-1}. 
\end{array} \label{eq17}
\end{equation} 
Such approach, however, should be used cautiosly. In \cite{chabert_2011}, for example, it is argued that there is a
resonance at the electron resonance-like frequency where the denominator in the expression for $C$ in Eq.~(\ref{eq17}) vanishes,
whereas, as argued previously, there is no such resonance in the electromagnetic dispersion relation of the first even mode.

The typical RF frequencies used to drive CCP discharges are well below the cutoff frequencies for the higher order even
modes ($n \ge 1$) (see Fig.~\ref{fig3}). Therefore, these modes are strongly damped in
the radial direction on the length scale of $s$ from their excitation location at the radial periphery inward. 
Indeed, following the same assumptions as in the derivation of Eqs.~(\ref{eq9-1}) and (\ref{eq9-2}), one obtains
\begin{equation}
\begin{array}{l}
K_{zs,n}^e \approx \frac{n\pi}{s}  \\
(K_{zp,n}^e)^2 \approx -\frac{1}{\delta_s^2} + \frac{n^2\pi^2}{s^2} \\
(K_{r,n}^e)^2 \approx k_0^2 -  \frac{n^2\pi^2}{s^2}
\end{array} \label{eq18}
\end{equation}
and 
\begin{equation}
\begin{array}{l}
K_{zp,n}^e \approx \frac{(2n-1)\pi}{2d}  \\
(K_{zs,n}^e)^2 \approx \frac{1}{\delta_s^2} + \frac{(2n-1)^2\pi^2}{4d^2} \\
(K_{r,n}^e)^2 \approx k_0^2 - \frac{1}{\delta_s^2} -  \frac{(2n-1)^2\pi^2}{4d^2}
\end{array}. \label{eq19}
\end{equation}
Therefore, these modes are significant only close to the radial plasma periphery
and are needed for satisfying boundary conditions at the sidewall (see \cite{lieberman_2002}).

\subsection*{Odd modes}
Following the same approach as for the even modes, wavevectors of the first odd mode in the long
radial wavelength limit are
\begin{equation}
\begin{array}{l}
(K_{r,0}^o)^2 \approx k_0^2 - \frac{1}{\epsilon_p sd} = k_0^2 + \frac{1}{\delta_t^2} \\
(K_{zs,0}^o)^2\approx \frac{1}{\epsilon_p sd} = -\frac{1}{\delta_t^2} \\
(K_{zp,0}^o)^2\approx k_0^2(\epsilon_p-1) + \frac{1}{\epsilon_p sd} = -\frac{1}{\delta_s^2} - \frac{1}{\delta_t^2}, \label{eq20}
\end{array}
\end{equation}
where $\delta_t \equiv \sqrt{-\epsilon_p sd}$ is used and the long radial wavelength limit was taken, so that
$|K_{zs,0}^o|s,|K_{zp,0}^o|d \ll 1$. This approximation is valid as long as $\delta_s$ and $\delta_t$ are large compared
to $l$. Note that $|\delta_t|$  provides a typical scale of the first odd
mode for sufficiently low $\omega$ (so that $k_0\ll |\delta_t|^{-1}$). In the collisionless limit $\delta_t \approx 
\omega_{pe}(sd)^{1/2}/\omega$ and the first odd mode propagates radially undamped. In the opposite
case of strong collisions ($\nu\gg \omega$), $\delta_t \approx (1+j)\omega_{pe}(sd/2\nu\omega)^{1/2}$,
consequently, $K^o_{r,0}$ is complex and the mode damps as it propagates from the radial periphery towards
the radial center of the discharge. The radial decay length of the first odd mode is usually smaller than that of
the first even mode, therefore the first odd mode is frequently too damped to form a standing wave.

As was done for the first even mode, one can find an equivalent circuit to describe propagation of the first odd mode.
Writing $K_{r,0}^o$ from Eq.~(\ref{eq20}) as
\begin{equation}
(K_{r,0}^o)^2 = -(j\omega\mu_0s/2+1/(j\omega\epsilon_0\epsilon_p2d))(2j\omega\epsilon_0/s) = -(j\omega L_s + Z_{sq})(2j\omega C_s), \label{eq21}
\end{equation}
where $L_s = \mu_0 s/2$ is the parallel sheath inductance and $Z_{sq} = 1/(j\omega\epsilon_0\epsilon_p 2d)$ 
is the lateral plasma impedance per square. Neglecting the small parallel sheet capacitance associated
with the lateral plasma impedance, $Z_{sq}=j\omega L_{sq} + R_{sq}$, where the sheet plasma dc resistance
$R_{sq}=1/(2d\sigma_{dc})$ with $\sigma_{dc} = n_e e^2/(m_e \nu)$ and sheet plasma inductance $L_{sq} = R_{sq}/\nu$.
It follows then that the appropriate circuit can be represented by a series distributed plasma sheet impedance and
sheath gap inductance along with a parallel capacitive sheath admittance (see Fig.~5). Note that for this 
equivalent circuit the current $I$ and $V$ denote different quantities than for the treatment of the first even mode.
Namely, here $I$ and $V$ represent the net radial sheet current in the plasma bulk and perturbation of the plasma potential
due to the first odd mode $V = - \int^0_{-l} E_{z,0}^o dz$ (provided the bottom electrode is grounded) \cite{sansonnens_2006}.

\begin{figure}[ht]
\centering
\epsfig{file=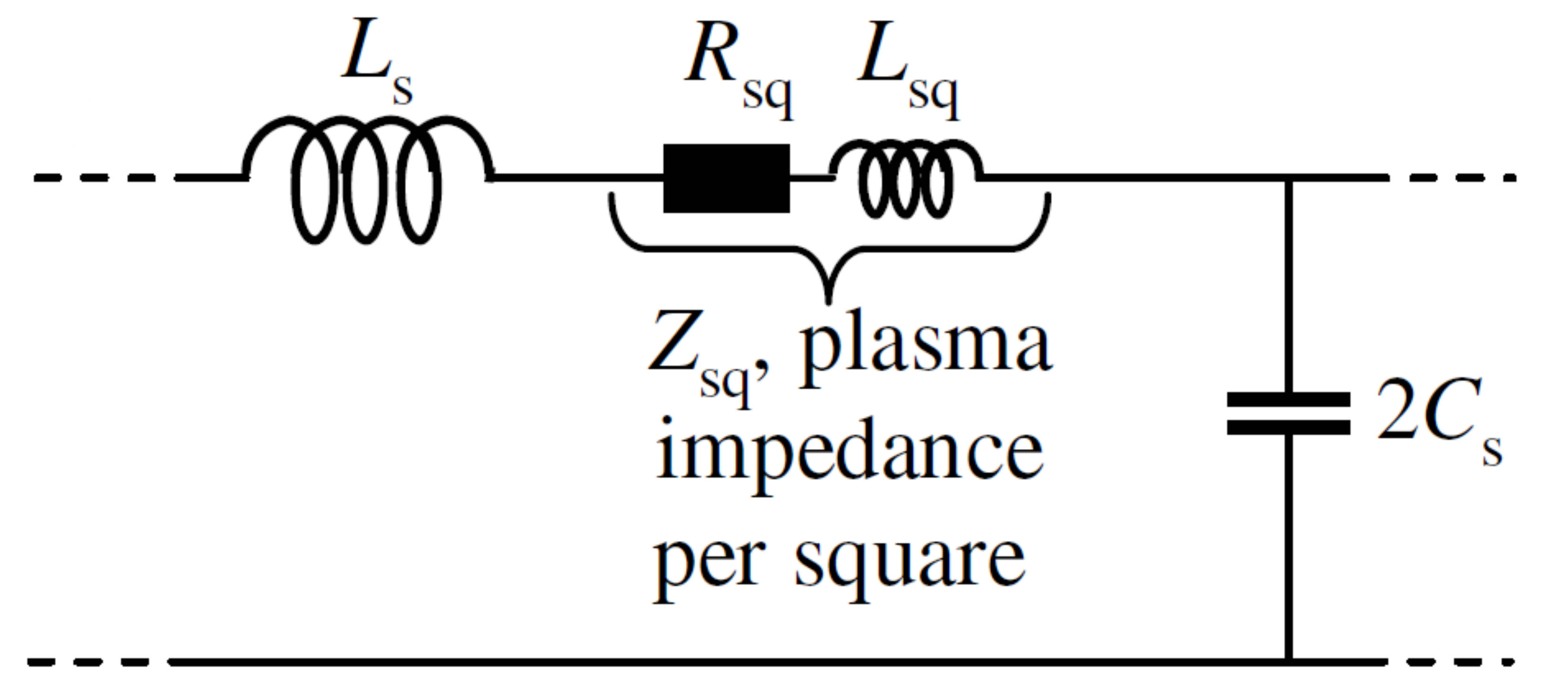,width=7cm}
\vspace{0.3cm}
\caption{Equivalent circuit for description of the first odd mode propagation, taken from \cite{sansonnens_2006}.}
\label{fig5}
\end{figure}

Similarly to the higher order even modes, the higher oder odd modes (with $n\ge 1$)
are also strongly damped in 
the radial direction on the length scale of $s$ from their excitation location at the radial periphery
towards the radial center. Following the same derivation that led to Eqs.~(\ref{eq9-5}) and (\ref{eq9-6}),
one obtains
\begin{equation}
\begin{array}{l}
K_{zs,n}^o \approx \frac{n\pi}{s}  \\
(K_{zp,n}^o)^2 \approx -\frac{1}{\delta_s^2} + \frac{n^2\pi^2}{s^2} \\
(K_{r,n}^o)^2 \approx k_0^2 -  \frac{n^2\pi^2}{s^2}
\end{array} \label{eq21}
\end{equation}
and 
\begin{equation}
\begin{array}{l}
K_{zp,n}^o \approx \frac{n\pi}{d}  \\
(K_{zs,n}^o)^2 \approx \frac{1}{\delta_s^2} + \frac{n^2\pi^2}{d^2} \\
(K_{r,n}^o)^2 \approx k_0^2 - \frac{1}{\delta_s^2} -  \frac{n^2\pi^2}{d^2} .
\end{array} \label{eq22}
\end{equation}

\section{The Standing wave, telegraph, skin, and edge effects}

In the literature there are four different effects mentioned in connection with the high frequency capacitively coupled plasmas:
the standing wave, telegraph, skin, and edge effects. 

As mentioned above, for the typical driving frequencies lying in the RF range only the first even and odd modes are excited in a CCP reactor, 
all other modes being strongly damped. The corresponding elecromagnetic waves propagate from the radial edge, where they are excited, 
towards the radial center. They are superposed with the waves propagating outwards from the center generated at the 
diametrally opposite point. Thereby they form standing waves mathematically described by the Bessel functions in Eqs.~(\ref{eq2}) and (\ref{eq3}). 
However, in the literature only the standing wave caused by the counterpropagating first even modes is called the 'standing wave effect'  \cite{schmitt_2002}, \cite{lieberman_2002}. 
This can be in part justified by the frequently observed stronger damping of the first odd surface wave mode propagating to the radial center from the radial edge 
compared to that of the first even mode, so that there is virtually no counterpropagating wave and thus technically no standing wave. 
This is however not the case in low-pressure weakly collisional CCP discharges driven by low or moderate RF frequencies.

The different axial electric field $E_z$ symmetries with respect to $z$ of the first even and odd standing wave modes cause different physics effects. 
The "the standing wave effect" consists in the first even mode (with $E_z(r,-z) = E_z(r,z)$) causing radial variations of the sheath voltages 
, which are symmetric in the both sheaths. The field symmetry of this mode is the same as that of the TEM mode excited in a CCP reactor
representing a vacuum capacitor in the abscence of plasma \cite{sansonnens_2006}. The characteristic signature of the first even mode is the decrease of the RF sheath voltage 
from discharge radial center towards its edge (mathematically, this is reflected by the behavior of $J_0$ Bessel function in Eqs.~(\ref{eq2})
and (\ref{eq3})). Due to the symmetry of the mode
the first even mode is excited by RF axial electric field,  which is symmetric with respect to the axial midplane. Since there is always such a field 
(it is needed to carry the discharge current in the sheaths by means of the displacement current and therefore is an essential part
of the discharge itself), 
the first even mode is always excited in the CCP discharges (as long as the excitation frequency $\omega<\omega_p/\sqrt{2}$). There are induced
radial currents propagating in the opposite directions at the top and bottom sheath - bulk plasma interfaces, so that this mode generates no net
radial current.

The first odd mode, in contrast, has $E_z(r,-z) = -E_z(r,z)$ and leads to the radial variations in the sheath voltage, which are also antisymmetric in the driven and grounded sheaths. 
The field symmetry of this mode is the same as that of the TEM mode excited in a coaxial cable (or a stripline waveguide) represented by 
a CCP reactor with high-conductivity plasma slab as the central conductor \cite{sansonnens_2006}.
Therefore, this mode leads to perturbations of the sheaths which are antisymmetric with respect to the axial midplane and do not result
in the perturbation of the discharge voltage, which is equal to the sum of the sheath voltages (provided the floating bulk plasma potential can
be neglected). This mode carries the RF plasma potential perturbation from the radial edge, provided it was excited there, in the direction of the radial center.
Due to the symmetry of the mode, it can be excited only in an asymmetric CCP reactor by the perturbation of the plasma potential
from its symmetric value arising due to the unequal capacitive division of the sheath voltages owing to the larger grounded sheath area
compared to the area of the driven sheath at the discharge radial edge. In this way, the first odd mode leads to increase
of the RF voltage amplitude on the driven electrode and to its decrease on the grounded electrode, respectively.
The equations describing radial propagation of the plasma potential perturbation
resemble those for the attenuated signal propagation in a lossy transmission line, hence the first odd mode and the radial uniformities
connected with it were dubbed 'telegraph effect' \cite{schmitt_2002}. This mode transfers net radial current in the plasma bulk and 
drives non-ambipolar currents in the sheaths close to the radial edge of the discharge and radial currents in the electrodes.
Whereas energy of the ions impinging on the conducting electrodes is radially uniform, it can become radialy nonuniform
on a dielectric substrate, which is potentially dangerous for the plasma processing \cite{howling_2004}, see also the next section.
The first odd (sometimes also called 'telegraph') mode might also dramatically affect the self-bias of the driven conducting electrode 
connected via a blocking capacitor to a power source \cite{howling_2004}.

As has been discussed previously, the skin effect comes into play when plasma densities are sufficiently high so that the skin depth becomes comparable to
the electrode gap. In this case the induced radial electric field becomes large, tends to be concentrated at the sheath-bulk plasma interfaces
and drives large radial currents there, which lead to screening of the axial electric and azimuthal magnetic fields from the radial center 
of the plasma towards its radial edge for the even mode. Another characteristic signature of strong skin effect is strong radial
electric field increasing from the radial center toward the edge (mathematically, this is reflected by the behavior of the $J_1$ Bessel
function in Eqs.~(\ref{eq2}) and (\ref{eq3}). The 'inductive' heating caused by the large induced radial electric field in this case
can prevail over the 'capacitive' heating caused by the axial electric field (\cite{lieberman_2002}, \cite{chabert_2005}). A similar situation
occurs in the inductively coupled plasma discharges, where such discharges are usually driven for low power in the capacitive regime and
for higher power in the inductive regime. It should be noted that although the skin effect changes the field pattern of the mode,
it does not change field symmetries, so that the mode identities are not altered by the skin effect.

Finally, the so-called edge effects are caused by the higher order even and odd modes, which are large close to the radial edge
on a distance of order of sheath width and can cause large field nonuniformities there \cite{lieberman_2002}.

\section{Non-uniformities of power absorption and the sheath voltages}

Uniformity of the power absorption and energy of the ions impinging on the substrate have
immediate relevance to the plasma processing. In this section we will consider such issues.

The fields in Eqs.~(\ref{eq2}) and (\ref{eq3}) depend on the plasma density $n(r)$ and plasma sheath width $s(r)$,
which in turn depend on the electromagnetic fields.  To model a CCP plasma discharge self-consistently, it is convenient
to use the equivalent circuit model, which provides a convenient representation of local coupling between the 
electromagnetic fields in the form of voltages and the currents based on the filed solutions of the Maxwell's equations
given in Eqs.~(\ref{eq2}) and (\ref{eq3}) along with the appropriate dispersion relations (see Section ???).
Using the transmission line equations with the obtained equivalent circuit elements it is possible to
describe propagation of the mode by calculating $V(r,t)$ and $I(r,t)$. Then, one can estimate the local
sheath width from the Child law using the obtained voltage $V$ and by calculating power absorbed in plasma
one can determine electron temperature and density from the particle and power balance equations,
respectively. By solving this self-consistent system of equations,  one can obtain the power absorption profile and estimate 
its influence on ion flux uniformity. Such a self-consistent approach for description of CCP discharges
featuring even mode (leading to the standing wave effect with a potentially strong the skin effect) 
was suggested in \cite{chabert_2004} and \cite{chabert_2005}. 

Following their approach (a part of it has led to Eq.~(\ref{eq17}) previously),
in order to calculate the absorbed power one has to augment that model by inclusion of the
resistive parts for the impedance $Z$ and the admittance $Y$, so that 
\begin{equation}
Z = R_{ind} + jL\omega, \label{eq23}
\end{equation}
and
\begin{equation}
Y^{-1} = R_{cap} + R_i + (jC\omega)^{-1}, \label{eq24}
\end{equation}
with $L$ and $C$ denoting the same quantities as in Eq.~(\ref{eq17}), 
$R_{ind}$ being the series resistance per unit length, $R_{cap} = R_{ohm} + R_{stoc} + R_{ohm,sh}$
being the parallel resistance per unit length, which comes from the ohmic heating resistance,
stochastic heating, and ohmic heating in the sheath (the latter two becoming significant 
at low pressure). 

$R_{ind}$ and $R_{ohm}$ can be calculated from the expressions for the power dissipation,
$2\pi r Re\left[\int^d_0 j_{\{r,z\}}(z) E^{*}_{\{r,z\}}(z) dz \right] = |I|^2 R_{\{ind.ohm\}}/2$, which yields
\begin{equation}
R_{ind} = Re\left\{ \frac{1}{2\pi r \sigma_{dc} \delta_s}\left[
\frac{\sinh(2d/\delta_s) - (2d/\delta_s)}{1+\cosh(2d/\delta_s)}\right] \right\} \label{eq25}
\end{equation}
with $\sigma_{dc} = e^2n_e/m_e\nu$ is the dc plasma conductivity (the
inductive stochastic heating \cite{turner_1993} can be incorporated using an effective
collision frequency \cite{lieberman_2005}), and
\begin{equation}
R_{ohm} = Re\left\{ \frac{\delta_s}{2\pi r \sigma_{dc} }\left[
\frac{\sinh(2d/\delta_s) + (2d/\delta_s)}{1+\cosh(2d/\delta_s)}\right] \right\}. \label{eq26}
\end{equation}
At low pressure, ohmic heating is dominated by stochastic
heating and ohmic heating in the sheath, which take place in the
sheaths and therefore are not influenced by the skin effect in the framework of the chosen
model with electron-free sheaths. Taking the corresponding quantities from \cite{chabert_2004}, one has
\begin{equation}
R_{stoc} = \frac{4 K_{stoc} (mT_e)^{1/2} s^2}{e^{1/2}\epsilon_0\pi r |V|}, \label{eq27}
\end{equation}
and
\begin{equation}
R_{ohm,sh} = \frac{2 K_{cap} K_{ohm,sh} m \nu s^3}{e\epsilon_0\pi r |V|}, \label{eq28}
\end{equation}
where $T_e$ is the electron temperature in eV, $K_{stoc} = 0.45$, $K_{ohm,sh} = 0.407$,
$K_{cap} = 1.23$ (the maximum sheath expansion is defined as $s_m = s K_{cap}$). 
Finally, the resistance per unit length due to power dissipation by ions flowing in the sheaths is \cite{chabert_2004}
\begin{equation}
R_i = \frac{4 K_v e h_l n_e u_B s^2}{\omega^2 \epsilon_0^2 \pi r |V|}, \label{eq29}
\end{equation}
where $K_v=0.83$, $u_B = (eT_e/M)^{1/2}$ is the Bohm velocity, $M$ is the ion mass,
$h_l = 0.86 (3 + d/\lambda_i)^{-1/2}$ is the axial plasma edge-to-center density radio and $\lambda_i$ is the ion
mean free path. 

To make the model self-consistent, one has to couple the transmission line equations
to the particle and power balance equations. In \cite{chabert_2007} two limiting
cases were considered: (i) non-local power deposition with the electron energy
relaxation length $\lambda_{\cal{E}} \approx \lambda (m_e/M)^{1/2}$ (with $\lambda$ the mean free path for
the $90^o$ scattering)  being larger than the discharge radius $R$, which is valid at low pressure and
(ii) local power deposition for the high pressure $\lambda_{\cal{E}} \ll R$.

Accordingly, assuming  global energy transport at low pressure and thus that
the electron density profile is defined by the ambipolar diffusion with the
uniform ionization profile, one can use the low-pressure diffusive solution proposed
in \cite{godyak_1986} 
\begin{equation}
n_e(r) = n_{e0} \left[ 1- (1-h^2_R)\frac{r^2}{R^2} \right]^{1/2} \label{eq30}
\end{equation}
with $h_R = 0.8 (4 + R/\lambda_i)$ the radial edge-to-center density ratio. 
Since the power balance at low pressure has global character, the absorbed power
for the power balance equation must be calculated by integrating it over the radial
domain,
\begin{equation}
P_{abs} = \frac{1}{2}\int\limits^R_0 R_{cap}\left|\frac{dI}{dr}\right|^2 dr
+ \frac{1}{2}\int\limits^R_0 R_{ind}|I|^2 dr. \label{eq31}
\end{equation}
The total power lost through the particle fluxes to the electrodes and the sidewall is \cite{chabert_2004}
\begin{equation}
P_{loss} = 2n_{e0}u_B(\pi R^2 h_l  + 2\pi R dh_R) \epsilon_T (T_e), \label{eq32}
\end{equation}
where $\epsilon_T(T_e)$ is the total energy loss per electron-ion pair created \cite{lieberman_2005}.
The electron temperature is determined from the particle balance
\begin{equation}
n_g K_{iz} \pi R^2 d = u_B(\pi R^2 h_l + 2\pi R d h_R), \label{eq33}
\end{equation}
where $K_{iz}$ is the ionization coefficient given in \cite{chabert_2004}.

In the other case of high pressure power deposition is local rather than global as in the case
of low pressure. Therefore, the electron density profile is determined from the local power balance,
where 
\begin{equation}
P_{abs}(r) = \frac{1}{2} R_{cap}\left|\frac{dI}{dr}\right|^2 + \frac{1}{2} R_{ind}|I|^2 \label{eq34}
\end{equation}
is the power absorbed at a radial position $r$ and 
\begin{equation}
P_{loss} = 4\pi r h_l n_e u_B \epsilon_T(T_e). \label{eq35}
\end{equation}
The local electron temperature in this case is to be found from the local particle
balance,
\begin{equation}
n_g K_{iz} d = h_l u_B. \label{eq36}
\end{equation}

The transmission line equations  $dV/dr = - ZI(r)$ and $dI/dr = -YV(r)$
for $V(r)$ and $I(r)$ have to be solved simultaneously with the
power balance equation $P_{abs}=P_{loss}$ to determine the electron density
and the particle balance equation to determine the electron temperature 
(taking the appropriate case in terms of the neutral background gas pressure)
either from global equations (\ref{eq31}), (\ref{eq32}), and (\ref{eq33})
in case of low pressure, or from local equations  (\ref{eq34}), (\ref{eq35}),
and (\ref{eq36}) in case of high pressure.

\begin{figure}[ht]
\centering
\epsfig{file=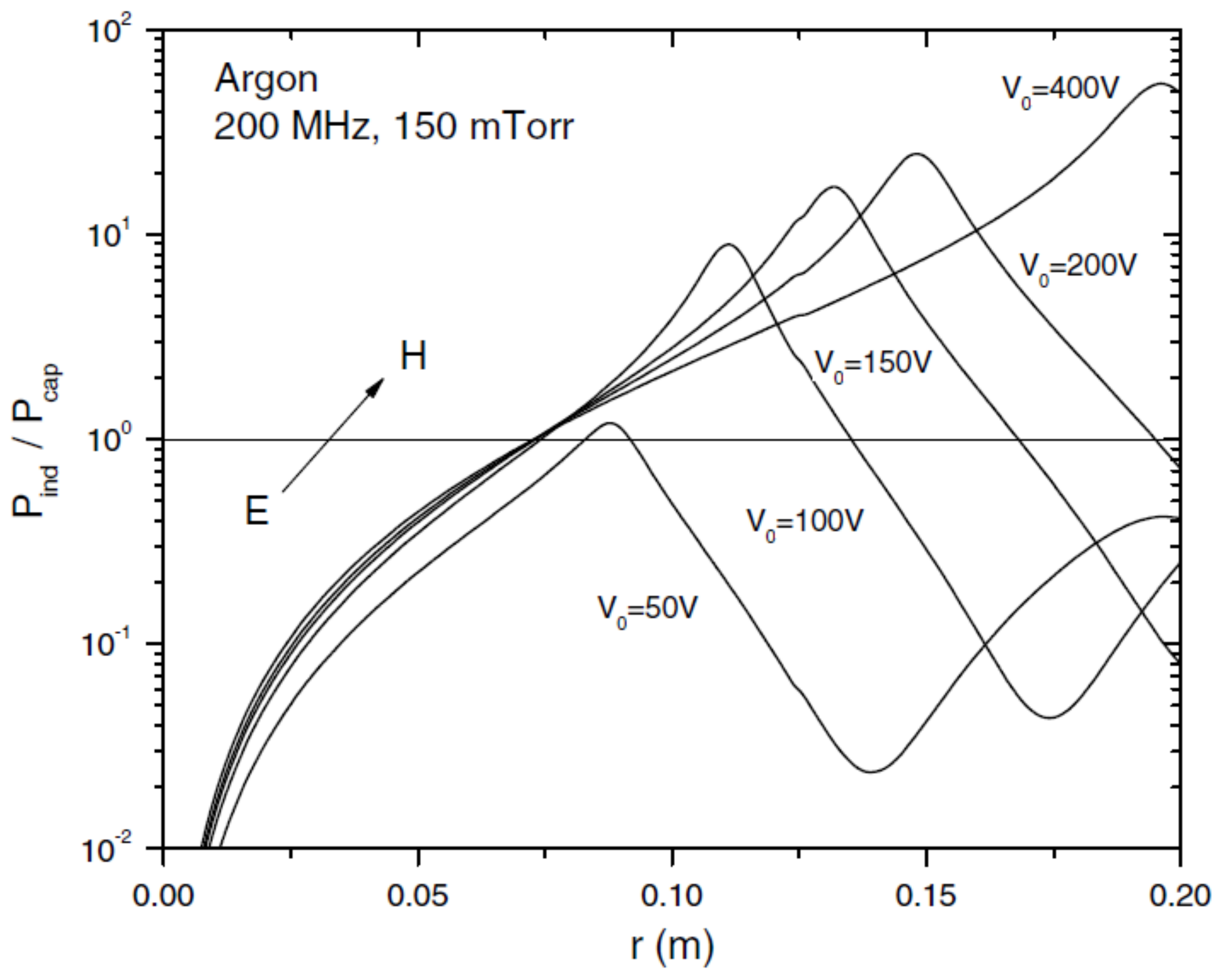,width=7cm}
\vspace{0.3cm}
\caption{Local E-H change of the heating regime in a CCP discharge illustrated by plotting ratio of the inductively deposited
power and capacitively deposited power $P_{ind}/P_{cap}$ versus $r$ for a discharge driven with 200 MHz voltage sources
with different central voltage amplitudes $V_0$ in the local regime under neutral gas pressure of 150 mTorr
, taken from \cite{chabert_2006}.}
\label{fig6}
\end{figure}

\begin{figure}[ht]
\centering
\epsfig{file=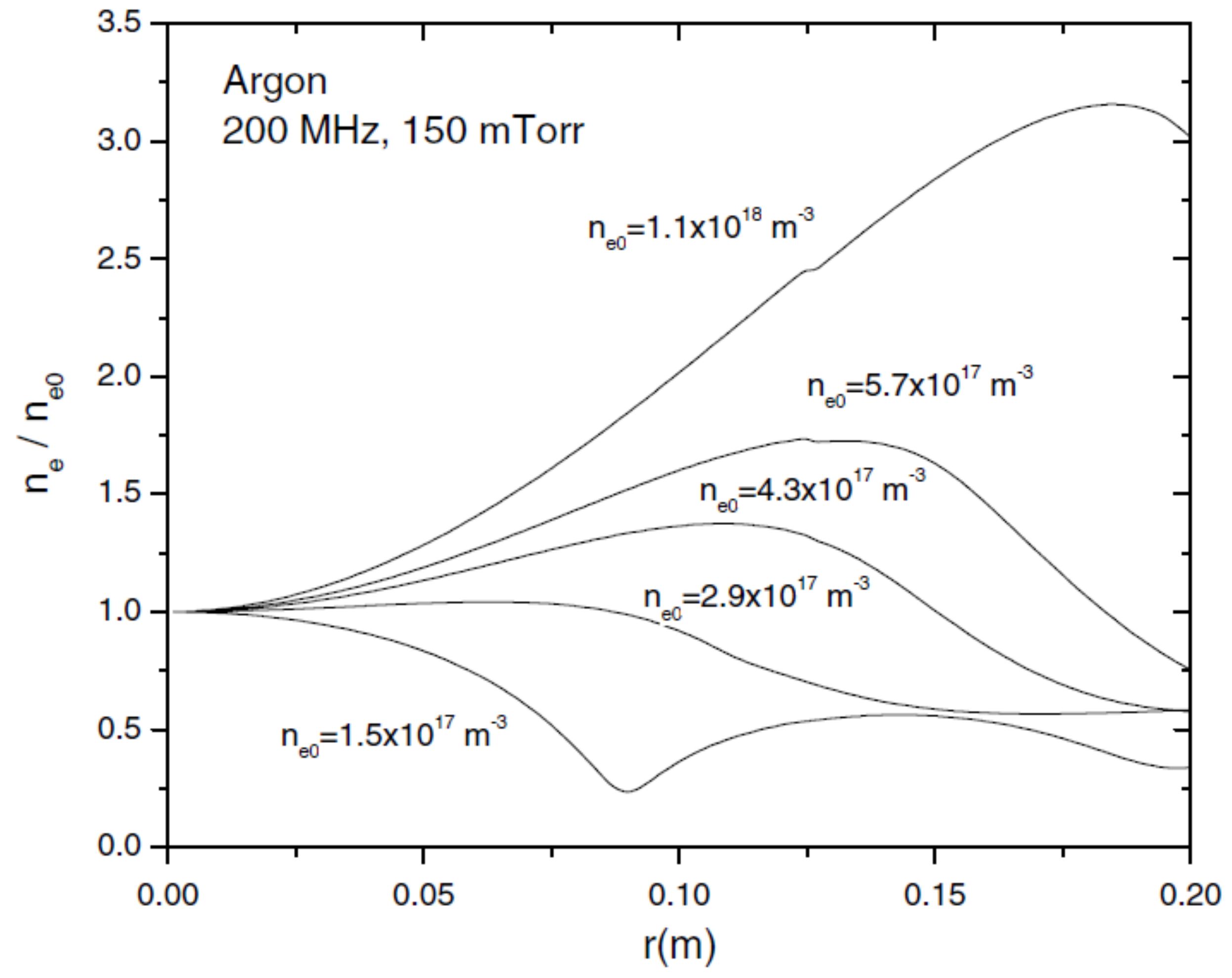,width=7cm}
\vspace{0.3cm}
\caption{Electron density profiles for the same cases shown in Fig.~\ref{fig6}
, taken from \cite{chabert_2006}.}
\label{fig7}
\end{figure}

The CCP discharges are normally heated in the capacitive regime (so that the first term on the
right hand side in Eq.~(\ref{eq31}) or Eq.~(\ref{eq34}) dominates over the second one
representing the inductive heating. In this regime the
standing wave dominates and the discharge voltage along with the deposited power profiles
show corresponding fall from the maximum at the center towards the radial edge. 
However, if one increases the power, the density is increased as well, the skin depth
becomes comparable to the electrode gap and the inductive heating wins over the capacitive
one and the discharge undergoes E-H regime change similar to the inductively coupled plasma
discharges (albeit with a reverse change of the heating regime with power increase compared
to the ICPs). In this regime the radial electric field is dominant and along with the 
power absorption profile shows typical pattern of the
skin effect, growing from the radial center towards the radial edge. If the case of low pressure
the E-H heating regime with increasing power occurs globally, in the case of high
pressure the heating regime change is local and, quite surprisingly, can occur in different directions
at several radial locations.
Fig.~\ref{fig6} shows results of the self-consistent transmission line 
model described above for CCP discharges driven in argon by RF voltage at 200 MHz and
150 mTorr neutral gas pressure with several central RF voltage amplitudes $V_0$ \cite{chabert_2006}. 
One can clearly see the transition from E to H heating regime and back occuring at several 
radial locations. The corresponding electron density profiles show the standing wave signature
of the fall from the radial center in the radial direction for the low power case ($V_0 = 50$ V) where
the capacitive heating regime dominates,
but  tend to have the skin effect signature of increase outwards from the center for higher powers,
see Fig.~\ref{fig7}.

The change in power absorption profile caused by the first odd mode (the telegraph effect) shows dramatically different pattern
from that caused by the first even mode (the standing wave effect). Owing to the symmetry of the odd mode the sheath voltage
of the sheath adjacent to the driven electrode is increased and the sheath voltage of the sheath adjacent to the grounded
electrode tends to be decreased by the same amount. It follows that while the power deposited in each of the individual sheath 
may be quite different from each other, but the total power is hardly changed \cite{howling_2004}. This might still  add
to the power deposition nonuniformity issue in the case of local power deposition (high pressure). 

Another issue important for the plasma processing is the radial uniformity of the peak ion energy for the ions hitting the
surfaces in a CCP reactor. This issue has been thoroughly investigated in \cite{howling_2004}. Although it might seem that 
the first even and odd electromagnetic modes excited in a CCP reactor should lead to the radial nonuniformity of the DC part 
of the plasma potential just as they do for its amplitude due to the standing wave and the telegraph effects
,  but this is normally not the case due to 
the high bulk plasma conductivity and small dc currents which flow through the bulk plasma. From the expression for
$\delta_t$ in the collisional case one can easily see that $\delta_t$ goes to infinity as the driving frequency vanishes,
which means that for the DC perturbations plasma potential must remain uniform.

As it was concluded in \cite{howling_2004}, the  uniformity of the ion peak energy (which is approximately equal to the 
DC component of the corresponding sheath voltage) strongly depends on the discharge set-up. Whereas the sheath voltage can be
uniform if the sheath touches the conducting electrode adjacent to it (in this case the electrode is held at a single
DC potential due to its high conductivity and the sheath voltage is also radially uniform), this ceases to be the case
on a dielectric substrate. The DC sheath voltage saturates at the self-rectifying value,
which is needed to ensure that period-averaged electron and ion currents are equal so the currents
are ambipolar \cite{garscadden_1962}. In case of the conducting electrode such self-rectifying value is attained in 
a global sense, so that the total DC ion and electron currents flowing through the electrode match each other
despite the local currents are nonambipolar (because of the electromagnetic effects the particle fluxes are 
nonuniform, while the sheah voltage must be uniform. This means that the current ambipolarity can be
achieved only globally and will inevitably be violated locally). The locally nonambipolar currents will
then be short-circuited on the electrode surface, which generates corresponding DC currents in the electrode.
In the case dielectric substrate the ambipolarity of currents will be achieved locally and the corresponding
self-rectifying sheath voltage will thus be radially nonuniform, which might be detrimental for plasma
processing. In the case of the conducting electrodes, it was also found in \cite{howling_2004} that
the self-bias of the driven electrode can be strongly changed by the telegraph effect compared
to value expected from the electrostatic theory. Since the self-rectifying sheath voltage is regulated
approximately by the largest RF amplitude in the sheath potential of the driven electrode
increases towards the radial edge due to the telegraph effect, its DC sheath voltage must increase
accordingly. This can only by achieved if the driven electrode has a negative self-bias. 

\section{Suppression of nonuniformities caused by the electromagnetic effects}

Since the processing plasmas are required to be as uniform as possible and all the electromagnetic effects
occuring in reactors with large electrodes driven by very high frequencies
described above lead to noniformities either in the ion fluxes to the substrate or the average 
energy of their directed motion to the substrate, 
it is a very interesting and important question of how to remove or alleviate such nonuniformities,
which is discussed in this section. Note that alternative concepts of high density plasma production
in CCP reactors without large area electrodes and/or VHF driving sources are not considered here. 

Because the standing wave and the skin effect lead to different radial power deposition profiles, 
one can assume that they can compensate each other under certain conditions. For example,
this can be seen in Fig.~13 in \cite{lieberman_2005}. However, such a compensation is hard
to control and it does not solve the issue of the sheath potential nonuniformity.

In \cite{sansonnens_2003} an elegant and efficient solution to ensure the uniformity of the axial field was suggested.
The Maxwell's equations do not allow a solution with purely axial field, they require a radial electric
field component. Assuming such a solution, one can ask if it is possible to generate such a solution
by a non-flat, shaped electrode. The resulting electric field must satisfy the boundary conditions
at each point of the electrode surface, which require that the tangential electric field vanishes so
that the electric field is normal at each point of the electrode surface. Thus, by enforcing 
the electrode surface to be perpendicular to the field solution at each point, one can calculate
the shape electrode, which turns out to be Gaussian \cite{sansonnens_2003}. The
plasma was to be confined by a dielectric slab and was assumed not to influence the
final solution. Later
analysis (\cite{chabert_2004_2}, \cite{sansonnens_2005}) has demonstrated this to be the case indeed
and generalized the method of calculating the electrode shape for the rectangular CCP discharges.
Such an approach tackles both the uniformity of the absorbed power and the sheath voltage issues.

In \cite{yang_2010} and \cite{yang_2010_2} the authors suggest two approaches to improve 
uniformity of the high-frequency, large-electrode-area CCP discharges, which also consist
in modifications of the driven electrode. The first approache consists in using graded conductivity electrodes (GCE),
i.e. to cover the driven electrode with a dielectic layer with conductivity decreasing from edge to the center.
As the RF wave propagates inwards from the edge
of the electrode, the penetration of the HF field into the dielectric increases. This increasing
penetration counteracts the increase in the electric field resulting from the standing
wave effect, and hence improves the uniformity of the resulting plasma. The second approach 
is to segment the driven electrode into separate sections and supply power to all the sections
ensuring the same phase or a constant phase shift between adjacent sections. 
The both approaches seem to yield a noticeable improvement
in the uniformity of the ion fluxes, but only modes improvement in the uniformity of the 
sheath voltages influencing the IEDF. 

The nonuniformities connected with the telegraph mode could be avoided by using symmetric
CCP reactors \cite{howling_2004}, in which such a mode is not excited.

\section{Experiments}

As the plasma processing industry goes in the direction of ever increasing electrode area and driving
frequency, the amount of corresponding experimental data is huge. Since a comprehensive
review of such experimental activities goes beyond focus of the present entry, only a few
experiments are mentioned here.

\begin{figure}[ht]
\centering
\epsfig{file=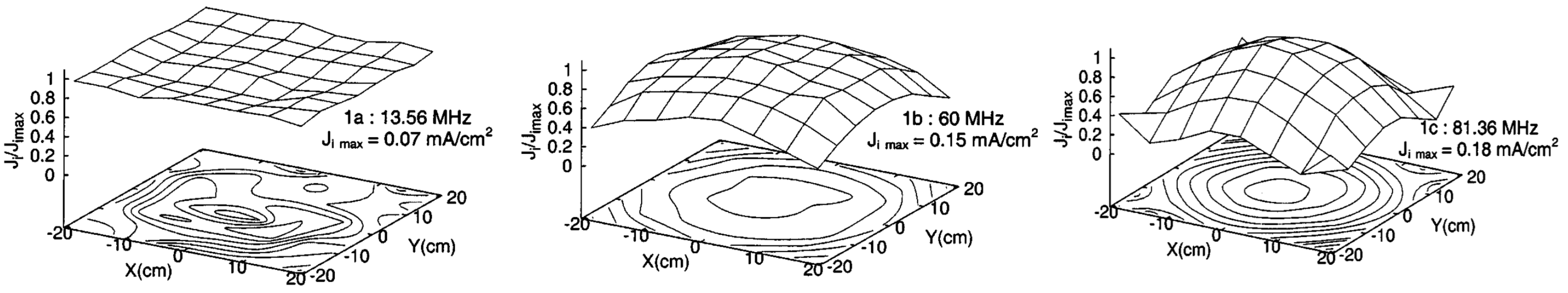,width=15cm}
\vspace{0.3cm}
\caption{2D ion flux uniformity at 150 mTorr and 50 W, for 13.56 MHz (a), 60 MHz (b), and 81.68 MHz (c). 
The standing-wave effect is observed at 60 MHz and is more pronounced at 81.36 MHz.
The figure is taken from \cite{perret_2003}.}
\label{fig8}
\end{figure}

One of the first direct experimental evidences of the electromagnetic effects causing nonuniformities of 
the ion fluxes was obtained through ion flux cartography in \cite{perret_2003}, see Fig.~\ref{fig8}. Here, the ion flux
cartography technique was used to measure the ion fluxes impinging on the electrode surface
by utilizing a 8 x 8 matrix of electrostatic negatively biased probes inserted into the grounded electrode.
The discharge was produced in argon gas at 150 mTorr pressure between two large-area square plates 40 cm x 40 cm 
separated by a distance of 4.5 cm, surrounded laterally by a 4 cm thick Teflon barrier. The measurements
were conducted for the same nominal power of 50 W at different driving frequencies.
Because of the relatively high pressure, the energy transport is local and the nonuniformity of the ion fluxes, which 
is a direct consequence of the nonuniform plasma density, is in this case connected with the nonuniformity
of the power absorption. One can
see that the ion flux is practically uniform for the frequency of 13,56 MHz, and it acquires 
a typical signature of the standing wave effect manifesting itself in the ion flux intensity falling off
from the discharge center to its edge. 
Further, nonuniformity of the discharge voltage due to the standing wave effect in a large area CCP discharge driven by a high frequency
source was demonstrated in \cite{sansonnens_1997} in the absence of plasma. The nonuniformity of the
plasma density due to the standing wave effect was experimentally revealed in \cite{chabert_2004_2}.
The suppression of nonuniformities caused by the standing wave effect through use of the
segmented electrodes with $180^o$ phase shift between the neighbor segments was checked
experimentally in \cite{monaghan_2011} with modest success.

\begin{figure}[ht]
\centering
\epsfig{file=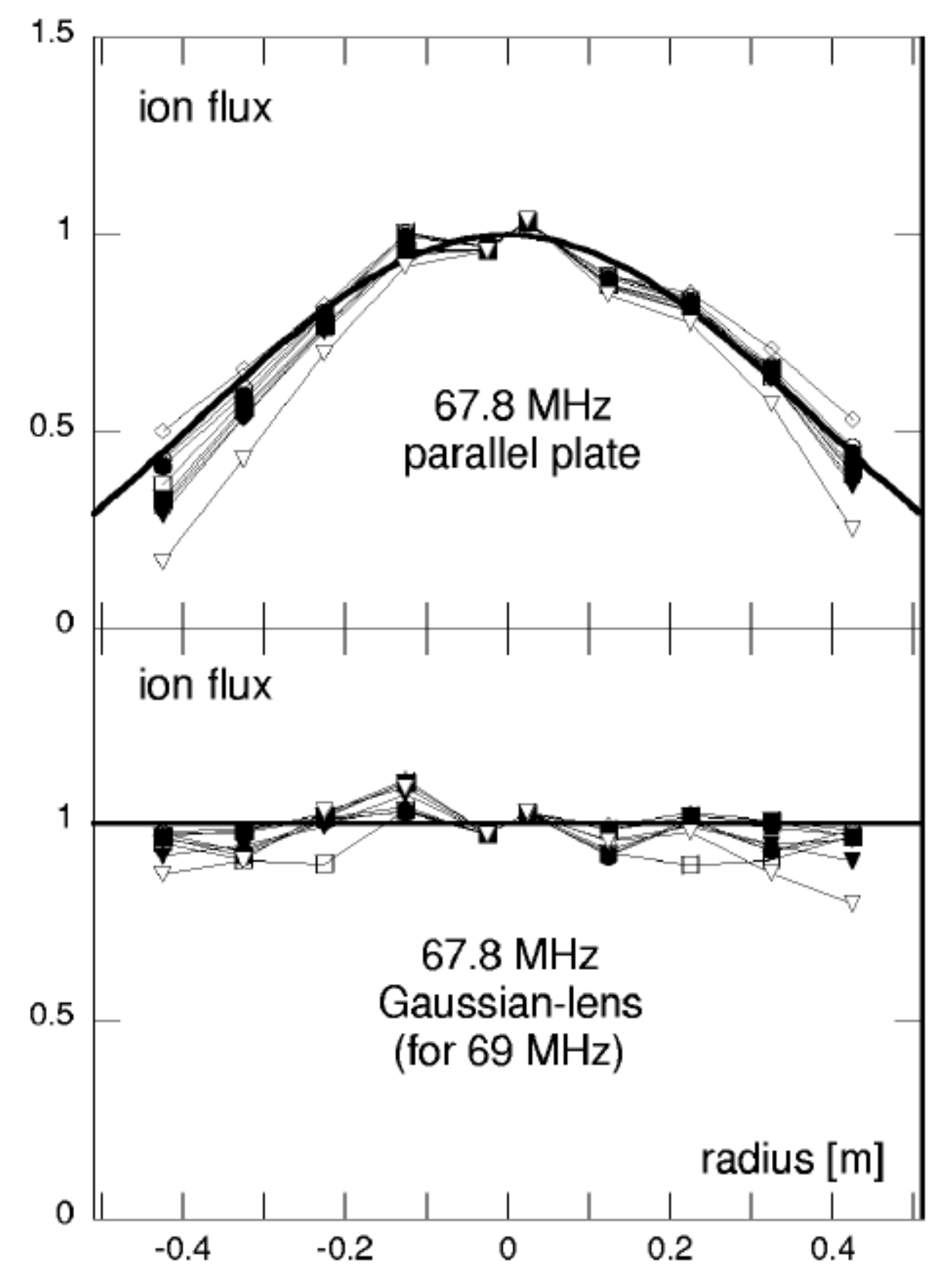,width=5cm}
\vspace{0.3cm}
\caption{Bottom: Measured radial profiles of the ion flux $I_{sat}(r)$ at 67.8 MHz
for the Gaussian-lens electrode filled with PTFE dielectric (design frequency
69.0 MHz). Top: the same measurements but with parallel plate electrodes.
Uniform profiles can be obtained using the lens electrode by suitable choice
of pressure and rf power, whereas the profiles for parallel plate electrodes
are strongly nonuniform and dominated by the standing-wave effect for all
plasma parameters indicated. Plasma parameters: 5 mTorr (100 and 300W),
10 mTorr (50 and 100 W), 25mTorr (50 and 100 W), 50 mTorr 50 W, 100
mTorr 50W, and 250 mTorr 200W. All values are normalized to the values
on axis. Error bars of $\pm 10\%$ are omitted for clarity. The figure is taken
from \cite{schmidt_2004}.
}
\label{fig9}
\end{figure}

The suppression of the standing wave effect due to the Gaussian lens electrode (see the previous section) was shown
experimentally in \cite{schmidt_2004}  by ion flux cartography and measuring the plasma optical
emission intensity. Fig.~\ref{fig9} taken from that work demonstrates that use of the Gaussian lens electrode 
designed for the appropriate driving frequency indeed leads to flattening of the ion flux radial profile.

An experimental observation of the induced radial electric field causing the skin effect in a CCP discharge was
made in \cite{ahn_2008}. For large densities, the inductive H heating mode was detected, with the
absorbed power profile increasing from the center to the the radial edge, a signature characteristic of
the skin effect.

The telegraph effect occuring in asymmetric CCP reactors leads to a radially nonuniform plasma potential,
which is due to the first odd mode generating antisymmetric perturbations in the ground and driven sheath
voltages. The nonuniformity of the plasma potential was demonstrated in \cite{howling_2005}, the 
asymmetric perturbations of the sheath voltages were observed indirectly through monitoring thickness
of the films deposited on the grounded and driven electrodes in \cite{howling_2007} (see Fig.~6 therein).
As mentioned above, the telegraph mode can cause redistribution of the currents at the electrodes
as a result of the self-rectification of the sheaths. If electrodes in a CCP reactor are exposed to 
plasma, the DC component of the sheath voltages, which determines
the average energy with which ions hit the electrode surfaces,
 is radially uniform (owing to uniformity of
the DC plasma potential and DC potential of the electrodes because of their high conductivity), which
was experimentally corroborated in \cite{perret_2005}. The corresponding DC currents flowing
radially in the electrodes to compensate for the nonuniform particle currents hitting the electrodes
as a result of the sheath self-rectification with the nonuniform voltages due to the standing wave
and the telegraph effects generating nonambipolar currents were directly measured in \cite{howling_2005_2}.
Another possible consequence of the telegraph effect when there is a delectric substrate covering
an electrode is that the surface charge and potential of the dielectric substrate can be negative and 
not only positive as for a uniform rf plasma potential, which was shown in \cite{howling_2001}.

%

\pagebreak

\pagebreak

\pagebreak

\end{document}